\documentclass[bibyear]{aa} 

\usepackage{amsmath,amssymb,natbib,subcaption}
\usepackage{txfonts}
\usepackage{graphicx,epsfig,epstopdf}
\newcommand{\haew}{H$\alpha$EW}
   \titlerunning{Unresolved versus resolved SSP}
   \authorrunning{H. Kuncarayakti et al.}
   
\begin{document}

   \title{Unresolved versus resolved: {testing the validity of} young simple stellar population models with VLT/MUSE observations of NGC~3603\footnote{Based on observations collected at the European Organisation for Astronomical Research in the Southern Hemisphere under ESO programme 60.A-9344.}}


   \author{
H. Kuncarayakti\inst{1,2}\thanks{\email{kuncarayakti@gmail.com}}, 
          L. Galbany\inst{1,2}, J. P. Anderson\inst{3}, T. Kr\"uhler\inst{3,4}, M. Hamuy\inst{2,1}
          }

   \institute{Millennium Institute of Astrophysics, Casilla 36-D, Santiago, Chile
        \and
             Departamento de Astronom\'ia, Universidad de Chile, Casilla 36-D, Santiago, Chile
             \and
             European Southern Observatory, Alonso de C\'ordova 3107, Vitacura, Santiago, Chile
             \and
             Max-Planck-Institut f\"ur Extraterrestrische Physik, Giessenbachstra{\ss}e, 85748, Garching, Germany
             }

   \date{Received XXXX; accepted 12 July 2016}

 
  \abstract
   {Stellar populations are the building blocks of galaxies including the Milky Way. The majority, if not all extragalactic studies are entangled with the use of stellar population models given the unresolved nature of their observation.
   Extragalactic systems contain multiple stellar populations with complex star formation histories. However, their study is {mainly} based upon the principles of simple stellar populations (SSP). Hence, it is critical to examine the validity of SSP models.
}
   {This work aims to empirically test the validity of SSP models. This is done by comparing SSP models against observations of spatially resolved young stellar population in the determination of its physical properties, i.e. age and metallicity.}
   {Integral field spectroscopy of a young stellar cluster in the Milky Way, NGC 3603, is used to study the properties of the cluster both as a resolved and unresolved stellar population. 
{The unresolved stellar population is analysed using the Ha equivalent width as an age indicator, and the ratio of strong emission lines to infer metallicity. In addition, spectral energy distribution (SED) fitting using STARLIGHT, is used to infer these properties from the {integrated spectrum}.}
Independently, the resolved stellar population is analysed using the color-magnitude diagram (CMD) for age and metallicity determination.   
   As the SSP model represents the unresolved stellar population, the derived age and metallicity are put to test whether they agree with those derived from resolved stars.
   }
   {The age and metallicity estimate of NGC 3603 derived from integrated spectroscopy are confirmed to be within the range of those derived from the CMD of the resolved stellar population, including other estimates found in the literature. The result from this pilot study supports the reliability of SSP models for studying {unresolved} young stellar populations.
   }
   {}


   \maketitle

%

\section{Introduction}
Stars together form stellar populations, which are the fundamental building blocks of galaxies in the Universe.
A simple stellar population (SSP) is usually defined as a group of stars distributed following an initial mass function (IMF), that was formed from a single cloud of gas with homogeneous chemical composition at the same time in one burst of star formation. As a result, stars within an SSP possess the same age and metallicity. Star clusters have been regarded as nature's closest approximation to the ideal SSP \citep[c.f.][]{bruzual10}.

Studies of external galaxies rely heavily on the SSP models. SSP models such as BC03 \citep{bc03}, Starburst99 \citep{leitherer99} and GALEV \citep{kotulla09}, among many others, are widely used as probes of unresolved stellar populations in galaxies. SSP models help interpreting the observed spectral energy distribution (SED) into meaningful physical quantities such as age, metallicity, luminosity, stellar content, and star formation rate.
{Correct determination of such properties is important for the study of galaxies itself as well as for other purposes.}
{For example}, in the studies of the hosts and environments of extragalactic transient objects such as supernovae (SNe) and gamma-ray bursts (GRBs), the properties of their stellar progenitors are derived from those of the underlying stellar populations \citep[see e.g.][]{galbany16a,galbany16b,galbany14,leloudas15,leloudas11,kruehler15,anderson15,hk13a,hk13b,levesque10}.
On that account, correct interpretation of the stellar population associated with the transient is necessary.

\begin{figure}
   \centering
   \includegraphics[width=\hsize]{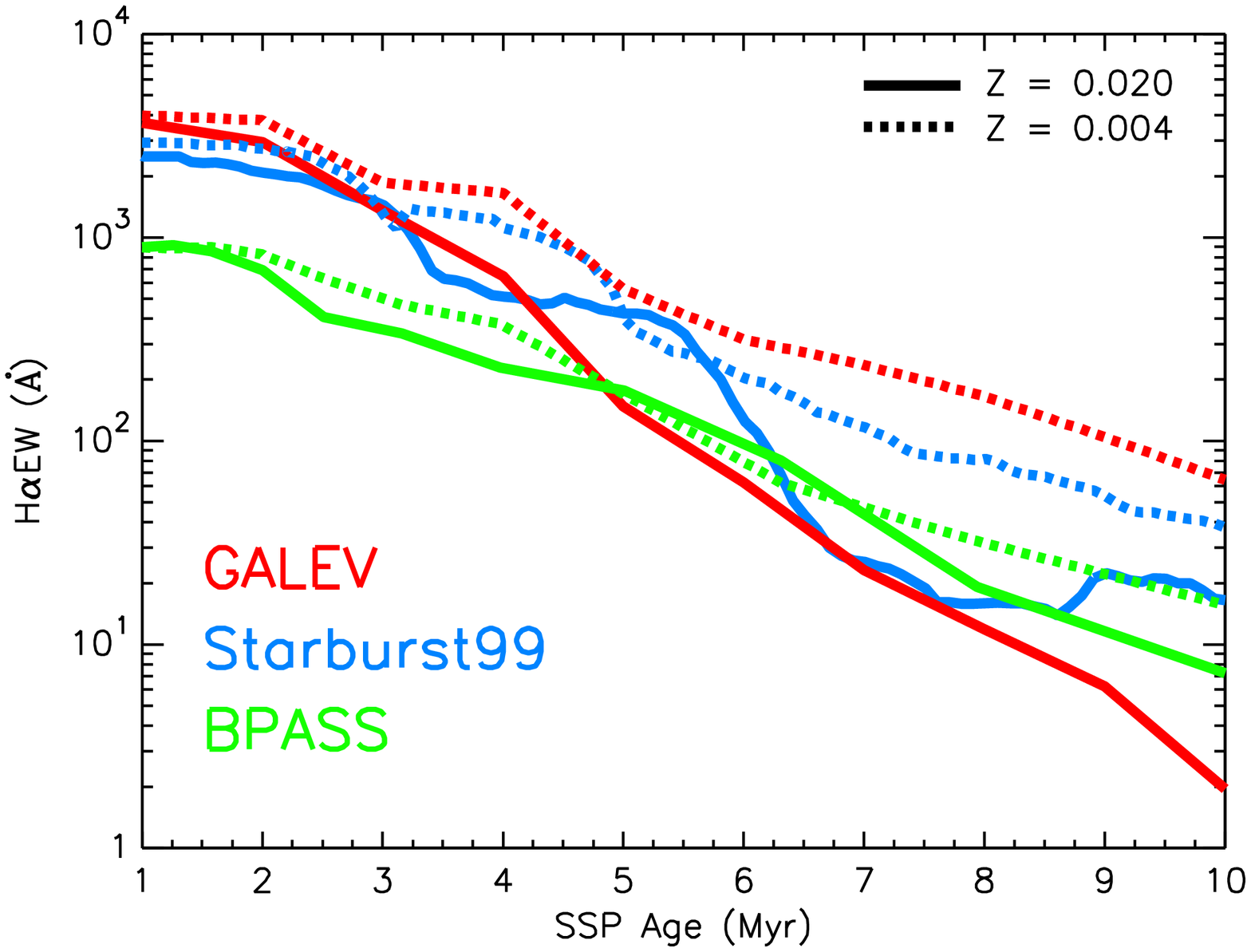}
      \caption{Comparison of H$\alpha$EW age indicator of BPASS {single-star} \citep{eldridge09}, GALEV \citep{kotulla09}, and Starburst99 \citep{leitherer99} {SSPs}, illustrating the differences between various SSP models. BPASS and GALEV H$\alpha$EWs were measured directly from the SSP model spectra (see section \ref{sec:emission} for the measurement method), while in Starburst99 the values are already given in tabulated form.
      All SSP models assume instantaneous star formation and a standard Salpeter IMF. 
      {As \haew$ $ also depends on the metallicity of the SSP model, two different metallicities are plotted to illustrate this effect. Solid lines indicate solar metallicity ($Z$ = 0.02) models and dashed lines indicate models with $Z$ = 0.004.}
              }
         \label{haewcomp}
   \end{figure}

While the validity of SSP models is of utmost importance, there are questions concerning the consistency between different SSP models and whether those are all well-{tested}. Different SSP models are generated using various different methods, thus the stellar population physical properties derived using one SSP model family may not necessarily be consistent with the result from another. 
{The main ingredients of SSP models usually consist of isochrones, a stellar spectral library, and an IMF \citep[see][for a review]{conroy13}. The choice of the isochrones and spectral libraries may be different from one model family to the other, and there can be variations at the more detailed levels such as the stellar evolution calculations for the isochrones and the selection of empirical or theoretical stellar spectra for the library, among others. These differences may eventually affect the constructed SSP models.}
Figure~\ref{haewcomp} shows the evolution of H$\alpha$ emission line equivalent width (EW) as an age indicator, according to three different SSP models families {(GALEV, Starburst99, and BPASS). These SSP models cover very young age ($\sim$Myr), and are thus suitable for the analysis presented here.}
It is apparent that despite the roughly similar behaviour in evolution, the H$\alpha$EW values may differ by a factor of two or more at a given age.

{To test the validity of SSP models in the analysis of complex extragalactic systems, a two-step procedure is required. First, the SSP models need to be compared to stellar clusters. Second, more complex systems are analysed using combinations of SSP models with non-instantaneous star formation, as one SSP model is considered too simplistic to represent such systems. In the current work, we aim to perform the first step in such a validity test. This first step serves as the basis for the subsequently more intricate step, which eventually will help in assessing whether the SSP-based techniques of extracting stellar population information in extragalactic systems are robust or not.}

Attempts to {test SSP models against stellar clusters} have been undertaken previously, although only for older stellar populations where the massive stars and ionized gas are no longer present \citep[see e.g.][]{renzini88}. \citet{beasley02} collected integrated spectra of 24 Large Magellanic Cloud star clusters and compared the spectroscopically-derived age and metallicity with those from the literature. In general, the age and metallicity derived using Lick/IDS spectral indices show consistency with the literature values, which were mostly derived from the resolved Color-Magnitude Diagram (CMD) or integrated colour for age and Ca-triplet spectroscopy for metallicity. Other efforts along the same line of study have also been done by \citet{koleva08} and \citet{riffel11}, who concluded that SSP models quite satisfactorily reproduce the observed parameters albeit there were minor caveats. {In the context of studying more complex objects with non-instantaneous star formation history,} more recently \citet{ruiz15} presented a comparison of resolved and unresolved stellar populations within a field in the Large Magellanic Cloud in a study of the star formation history in the region.

On the other hand, SSP models corresponding to young stellar populations ($\sim$Myr age) are less well-established compared to the older population ($\sim$Gyr) grids {\citep{conroy13}}. {Many SSP models do not include young grids or take into account nebular emissions coming from young stellar populations, while they commonly provide grids for old populations.}
These young stellar populations are dynamic objects associated with active star-forming regions where the interplay between the ionizing massive stars and the surrounding gas is intense, and energetic events such as core-collapse SNe and long-GRBs occur {\citep[see e.g.][]{portegies10}}. Even in a galaxy preeminently consisting of old stellar populations, the young populations may dominate the emissions in the ultraviolet/optical regime thus affecting significantly the observed integrated properties {\citep{meurer95,anders03}}.


In this paper we report our pilot study {examining the reliability of} young SSP models, by using VLT/MUSE integral field spectroscopy (IFS) of a well-studied young stellar population, NGC 3603. It has been suggested that the cluster is as young as a few Myr \citep[e.g.][]{melnick89,sagar01} and was formed in an instantaneous burst of star formation \citep[][]{kudryavtseva12,fukui14}, {and} thus represents a good approximation for a young SSP.
As young stellar populations are still associated with ionized gas, IFS offers the most efficient way to obtain the integrated spectrum of the cluster, including the ionized gas, in addition to the individual spectra of the member stars. 
{With this technique, it is possible to obtain simultaneously the integrated and resolved information of a stellar population from a single dataset.}
This would have taken enormous time to conduct with conventional slit or fiber spectroscopy, and to our knowledge this is the first of such efforts in {testing} young ($\sim$Myr) SSP models. 

The paper is organized as follows. Following the introduction, the observations and data reduction are described in Section 2, then the analysis of the integrated spectrum for age and metallicity estimates is presented in Section 3. In Section 4 we analyse the properties of the resolved stars, and compare these to the unresolved stellar population from the integrated light. Section 5 provides a final summary.

\section{Observations and data reduction}

\begin{figure*}[]
   \centering
   \includegraphics[width=\hsize]{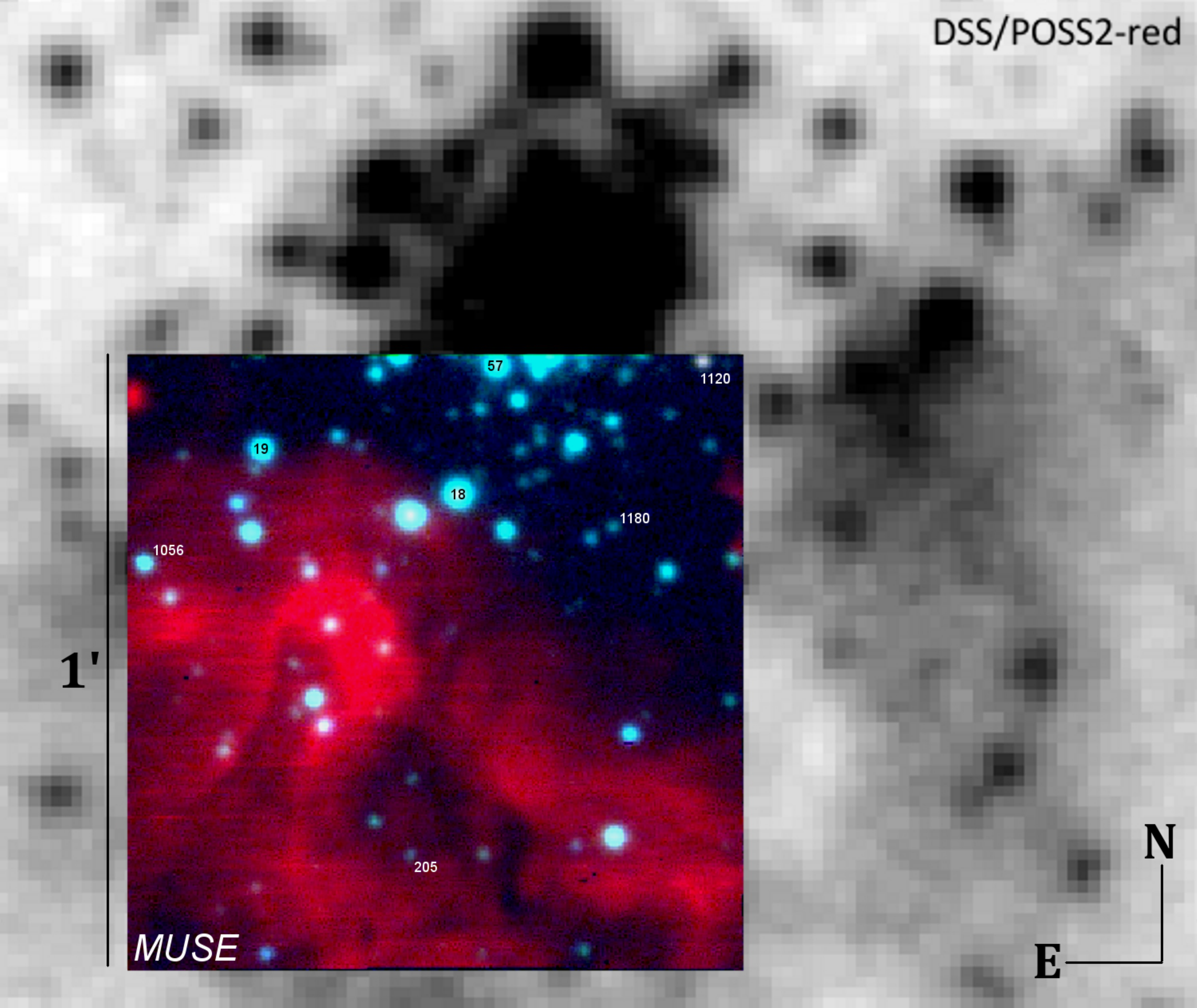}
      \caption{Pseudo-RGB colour image of the MUSE field of view (FoV) of NGC 3603 generated from the datacube, superposed on a POSS2-red image from the Digitized Sky Survey (DSS). North is up and east is to the left. Each side of MUSE FoV is approximately 1 arcmin (corresponding to 2 pc at the distance of 7 kpc). The core of the cluster lies just outside the field of view to the north-north-west direction, and was intentionally missed in the observations to avoid the severe crowding.
      Red colour corresponds to H$\alpha$ and blue/green correspond to the continuum at 4700 and 5500 \AA, respectively.
      {The stars whose spectra are shown in Figure \ref{spec} are indicated with their identification numbers, based on the WEBDA catalogue.}
              }
         \label{figrgb}
   \end{figure*}

The observations were conducted using the Multi-Unit Spectroscopic Explorer (MUSE) integral field spectrograph \citep{bacon14} of the Very Large Telescope (VLT) atop Cerro Paranal Observatory, Chile. As part of the MUSE Science Verification run, NGC 3603 was observed on 2014 June 29 (Programme 60.A-9344, PI: Kuncarayakti\footnote{Shared observation programme with PIs Vink and Gonz\'alez-Fern\'andez as in the MUSE Science Verification run similar proposals were combined together for efficiency.}). 
MUSE was utilized in the Wide Field Mode, giving a 1 arcmin$^2$ square field of view sampled by 0.2"$\times$0.2" spaxels across 4650--9300 \AA. The mean spectral resolution is around 3000, with dispersion of 1.25 \AA/pixel. The observations were done in non-photometric conditions under $\sim$0.8" seeing.  The raw data were reduced using the MUSE data reduction pipeline integrated within the Reflex environment \citep{freudling13}, which includes procedures of bias subtraction, flatfielding, subtraction of sky from blank sky pointings, wavelength and flux calibrations, spectral extraction and datacube reconstruction. The resulting datacube from a single 300 s integration was used in this work. Figure \ref{figrgb} shows a pseudo-RGB image of the whole MUSE field of view of NGC 3603. The image quality was FWHM $\approx$ 0.8" as measured in the collapsed datacube, consistent with the seeing monitor, demonstrating the excellent performance of MUSE.

\begin{figure*}
   \centering
   \includegraphics[width=\hsize]{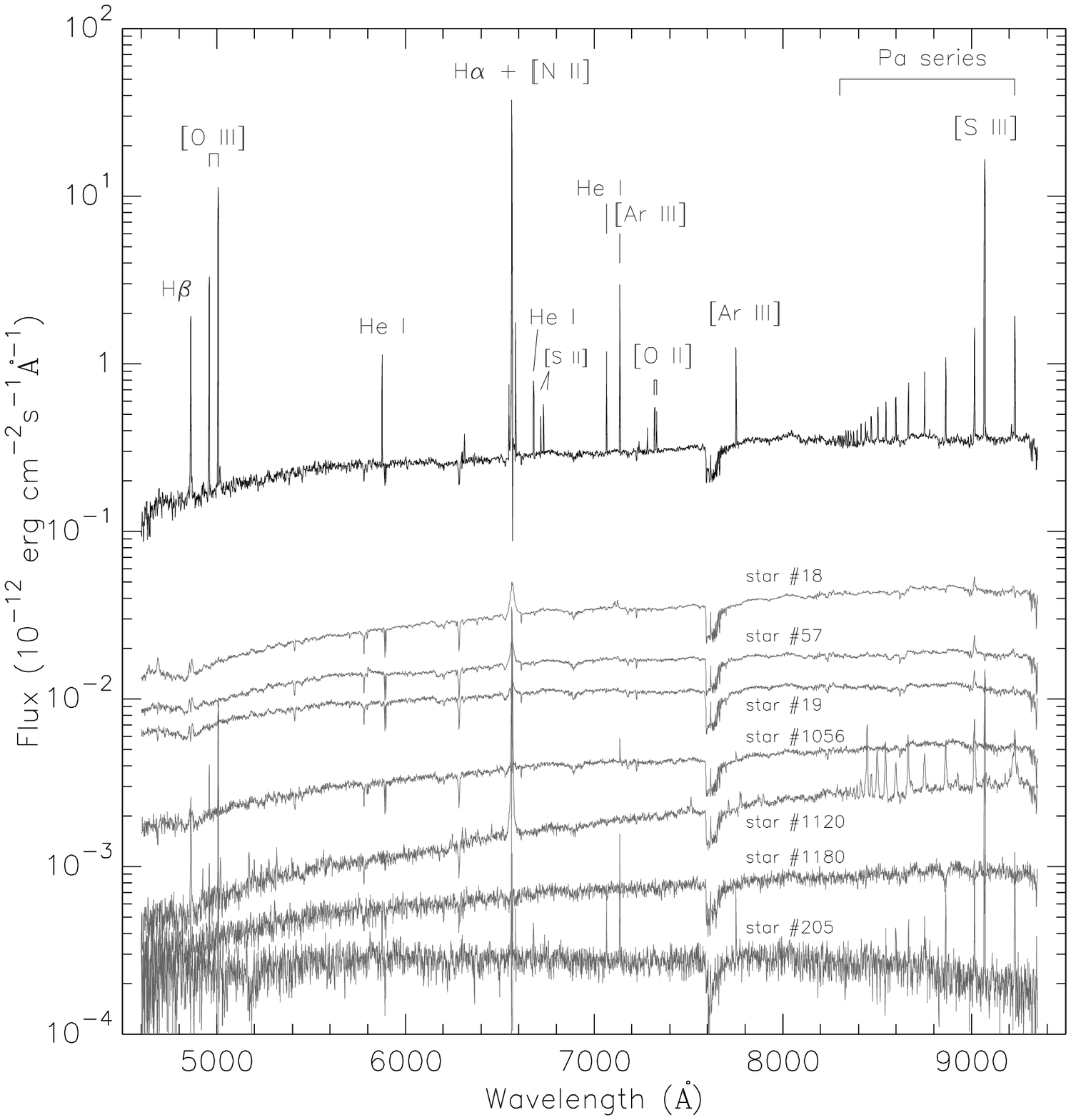}
      \caption{The observed integrated spectrum of NGC 3603 within the whole MUSE field of view (top; solid black line), plotted alongside a montage of spectra from several individual stars in the field (solid grey lines). 
{The individual stars are identified with their respective identification numbers (see Figure~\ref{figrgb}).}      
      The integrated spectrum is the sum of all the spaxels in the field of view, {and} thus includes all the individual stellar spectra and the associated nebulosity.
      The strongest emission lines in the integrated spectrum are indicated. Spectra are wavelength- and flux-calibrated.
              }
         \label{spec}
   \end{figure*}

The datacube was subsequently analyzed using the QFitsView\footnote{\url{http://www.mpe.mpg.de/~ott/dpuser/qfitsview.html}} visualization software \citep{ott12} and \textsc{Iraf}\footnote{\textsc{Iraf} is distributed by the National Optical Astronomy Observatory, which is operated by the Association of Universities for Research in Astronomy (AURA) under cooperative agreement with the National Science Foundation.}.  
We extracted the spectrum over the full MUSE field of view, integrating at the same time both the spectra of individual stars, together with the interstellar diffuse light.
Figure~\ref{spec} shows the integrated spectrum of NGC 3603, exhibiting a multitude of strong emission lines emitted by the ionized gas superposed on a continuum originating from the stellar component. A montage of several individual stellar spectra is also shown in the figure.
The individual stellar spectra were extracted using an aperture {in comparable size to the point spread function FWHM, i.e.,} 4 spaxels (0.8") radius. These integrated and individual spectra were then used in the subsequent analyses. \textit{VRI} magnitudes of the individual stars were also calculated by applying synthetic photometry to the individual star's spectra. These were then cross-matched and compared with the entries at the WEBDA\footnote{\url{http://www.univie.ac.at/webda/}} online stellar clusters database. As our observations were done under non-photometric conditions, there were zero point offsets between our synthetic \textit{VRI} magnitudes with the photometry available at WEBDA, which were taken from the work of \citet{sagar01}. We calculated the mean offsets for each filter and applied them to the synthetic magnitudes of each star.

\section{Analysis of the integrated spectrum}

The integrated spectrum of NGC 3603 is considered analogous to that of an unresolved stellar population, observed spectroscopically. The spectrum contains both the continuum emission contributed by the individual stars and the nebular component emitted by the surrounding ionized gas.

The age and metallicity of the cluster were derived using two independent methods that both use SSP models. The first method utilizes the emission lines; it compares the equivalent width of the H$\alpha$ emission line with SSP models to determine the age at a fixed metallicity, the latter derived via strong-line method. The second method fits the continuum of the integrated spectrum with the spectral energy distributions of SSP models to simultaneously determine the age and metallicity.

\subsection{Emission line analysis}
\label{sec:emission}

The equivalent width of {the} H$\alpha$ emission line (\haew) is used to derive the stellar population age. As it measures the flux ratio between the line emission from the ionized gas and the continuum light emitted by the stellar component,  it is proportional to the number ratio of massive, ionizing OB stars compared to the lower mass, non-ionizing stars. Thus in an SSP it is expected that the \haew~decreases over time as the short-lived massive stars die first even though all the SSP stars possess the same age (see Figure~\ref{haewcomp}). The use of \haew~as good age indicators for young stellar populations has been demonstrated by e.g. \citet{reines10,leloudas11,hk13a,hk13b}.

From the integrated MUSE spectrum of NGC 3603 we measured the \haew, as well as gas-phase metallicity in terms of oxygen abundance. The abundance was derived first as the evolution models of \haew~depend on the assumed metallicity {(see Figure \ref{haewcomp})}.
We use the strong-line method using the N2 \citep{storchi94} and O3N2 \citep{alloin79} indices calibrated according to \citet{marino13}. 
Using the task \texttt{splot} within the \texttt{onedspec} package in \textsc{Iraf}, the emission line fluxes were measured for metallicity determination.
The derived metallicity for NGC 3603 from the N2 index is 12 + log(O/H) = 8.09 dex, while it is 12 + log(O/H) = 8.07 dex from the O3N2 index. The averaged metallicity therefore corresponds to {$Z$ = 0.0034}, assuming solar oxygen abundance of 12 + log(O/H)$_\odot$ = 8.69 dex or Z$_\odot$ = 0.014 \citep{asplund09}. 
{For the subsequent analysis, SSP models with nearest metallicity value of $Z = 0.004$ are used.}


Also using \texttt{splot}, the \haew~of NGC 3603 was measured as $378 \pm 0.3$ \AA. The same method of measurement was also done for GALEV and BPASS SSP spectra, resulting in the \haew~evolution shown in Figure~\ref{haewcomp}.
This observed \haew~of NGC 3603 was then compared to the Starburst99 SSP models with $Z$ = 0.004 and {a} standard Salpeter IMF ($\alpha = 2.35$, $M_{\textrm{up}} = 100$ $M_\odot$), yielding {an} age of 5.05 Myr. As a cross check using ROBOSPECT, an automatic spectral line equivalent width measurement software \citep{waters13}, a value of \haew~= 355 \AA~was derived, corresponding to the similar age of 5.09 Myr with the same Starburst99 SSP. 

{Using the \haew$ $ value from \texttt{splot} measurement, age estimates were also derived using GALEV and BPASS SSP models. 
With models of the same $Z = 0.004$ metallicity, the resulting age estimates were 5.74 and 3.94 Myr, for GALEV and BPASS respectively. 
Although the results of the three SSP model families are consistent at the value of around 4-6 Myr, in line with Figure~\ref{haewcomp} this illustrates the different outcomes in estimating stellar population age when using different sets of SSP models.}


\subsection{SED fitting}

We further analysed the integrated spectrum of NGC 3603 using the STARLIGHT code \citep{cid05} to estimate the age and metallicity. STARLIGHT utilizes the method of fitting the observed SED with a library of SSP model spectra with different ages and metallicities, and returns the best combination of SSP spectra that matches the object SED. 
The fitting was done to the continuum and absorption lines while the emission lines were masked out. The fitting process is similar to the one described in \citet{galbany14}{, only with different wavelength range due to MUSE not covering the blue part of the optical spectrum. We used the spectral range from 4650 \AA$ $ for the fit, and experimented with different red wavelength cutoffs. Fitting the whole MUSE spectral range up to 9300 \AA$ $ generally increases contamination from old stellar populations, and the fitting was found to be optimal (i.e., dominated by young population) when limited to around 7000 \AA.
}

\vspace{2cm}

The spectral library used by our STARLIGHT fit consists of 66 SSP spectra, each with different ages and metallicity. These were modified from \citet{bc03} SSP models \citep[see][]{bruzual07}, using MILES spectral library \citep{sanchez06},  {a} Chabrier IMF, {the} Padova 1994 evolutionary tracks, and TP-AGB star recipes by \citet{marigo07} and \citet{marigo08}.

\begin{figure}[h]
   \centering
   \includegraphics[width=\hsize]{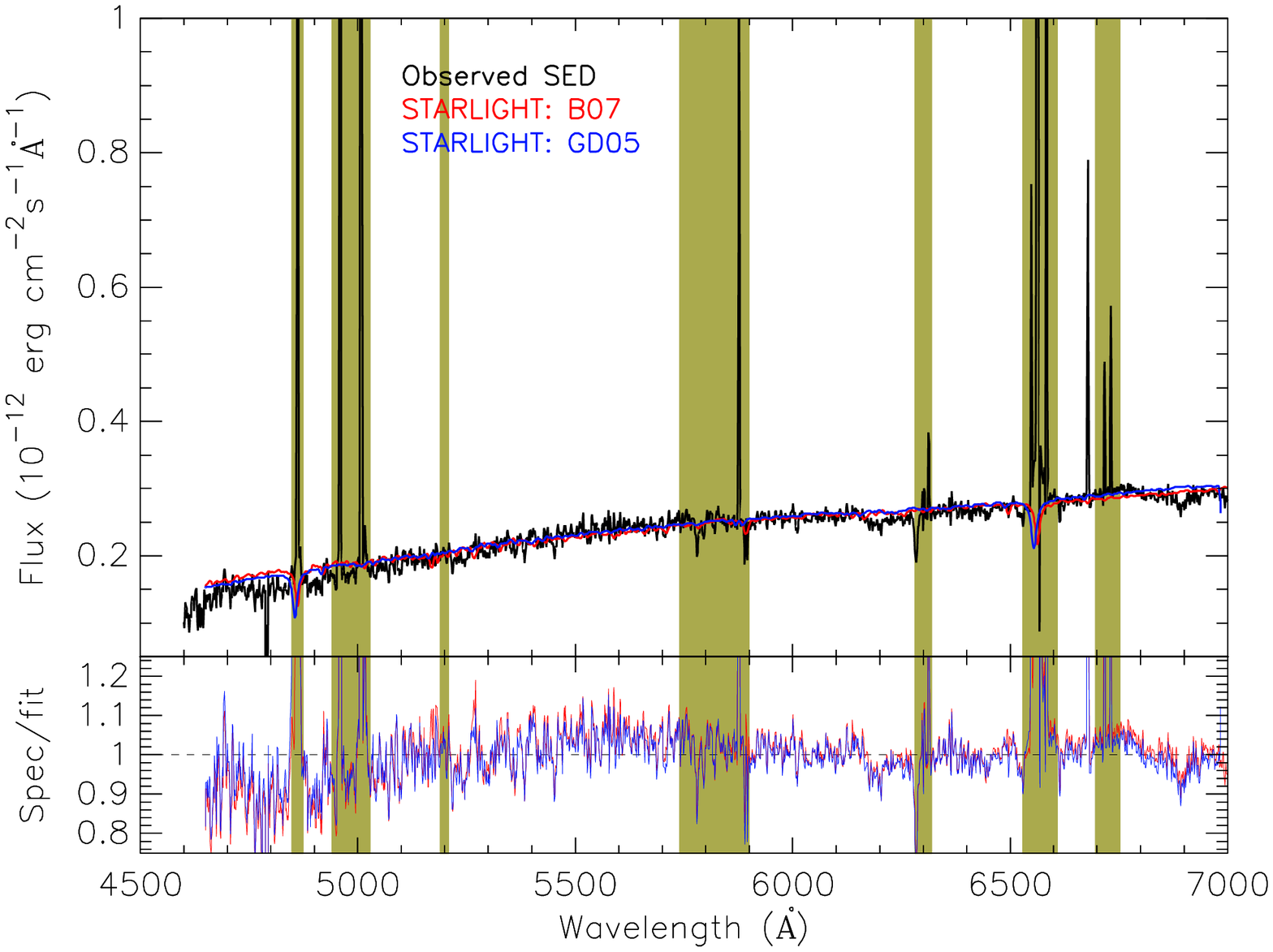}
      \caption{
     { The result of the STARLIGHT fit using the SSP models from \citet[][red]{bruzual07} and \citet[][blue]{gonzalez05} compared to the integrated spectrum of NGC 3603 (black). Masked regions are shaded in {dark} yellow. Fit residuals are shown on the bottom panel. The horizontal dashed line indicates the ratio of unity between the observed spectrum and model fit.}
              }
         \label{sedfit}
   \end{figure}

From the SED fitting, STARLIGHT found that the integrated spectrum of NGC 3603 is best matched by an SSP with log age of 6.61 ($\approx$ 4.1 Myr) and metallicity $Z$ = 0.008, additionally with negligible contribution from older SSPs of log age $\approx8$ and 10. This 4 Myr estimate agrees well with the {4-6} Myr result from the \haew~method.
{For a comparison, we also tried to use a different SSP model for the fitting \citep{gonzalez05}. The best fit contains stellar population with age of 3.0 Myr, although with significantly higher metallicity of $Z$ = 0.03 and heavier contamination from the old populations in which they became the dominant component. Therefore, the result using \citet{bruzual07} SSP models is preferred.
Figure~\ref{sedfit} shows the STARLIGHT fits compared to the observed integrated spectrum of NGC 3603. It is interesting to note that although the \citet{bruzual07} and \citet{gonzalez05} SSP models produce very similar fits, the age/metallicity solutions are markedly different.
}


%
%


\section{Comparison with resolved stellar population}

With the age and metallicity of NGC 3603 derived from the integrated spectrum, we now consider the estimates determined from the {individual stars that are part of} NGC 3603. 
As resolved stars are considered to be the canonical form of stellar populations, such comparison enables the assessment on the reliability of the SSP models.
If the SSP models are correct, the properties derived from integrated spectrum should match with those derived from the resolved stellar population.

In most cases in the literature the age of NGC 3603 was derived using the color-magnitude diagram (CMD), a common method to analyse stellar clusters in which the photometry of the individual stars is compared to theoretical isochrones in order to derive simultaneously the cluster's age, metallicity, distance and reddening. 
{Previously, it has been generally found} that NGC 3603 contains a very young stellar population with age of a few Myr and metallicity around solar or lower, although there is no single consensus as to what is the exact age, and an apparent lack of discussion on the metallicity.

\citet{melnick89} presented \textit{UBV} CCD photometry of the cluster, and from the CMD {inferred} an age between 2 and 3 Myr, with 2 Myr spread. They supplemented their analysis with an estimate of metallicity using slit spectroscopy of the associated nebula, which yielded 12 + log(O/H) = 8.39. Subsequently, utilizing \textit{UBVRI} photometry \citet{pandey00} concluded a mean cluster age of $\lesssim 1$ Myr with an age spread of a few Myr within the cluster, and \citet{sagar01} found an age of $3 \pm 2$ Myr. 
{\citet{sung04} resolved the inner core and studied the stellar content of the cluster by utilizing \textit{Hubble Space Telescope (HST)} Wide Field and Planetary Camera 2 (WFPC2) observations. They derived age of $1 \pm 1$ Myr for the inner part of the cluster, and $\lesssim 5$ Myr for the outer parts.
{Prior to the \textit{HST}, well-resolved observation of the inner part of the cluster was not available.}
Later, \citet{melena08} used \textit{HST} Advanced Camera for Surveys (ACS) images to corroborate this result, finding that the most massive stars (some reach $\gtrsim120$ $M_\odot$) are coeval within 1-2 Myr but there could be an age spread up to 4 Myr in the cluster. Utilizing ground-based high spatial resolution adaptive optics observations in the near-infrared, \citet{harayama08} derived an age of $\lesssim 2.5$ Myr for the cluster.
}

Some estimates using the lower part of the CMD that contains the pre-main sequence (PMS) stars may suggest different ages.
\citet{beccari10} derived an age spread between 1-20 Myr assuming solar metallicity, Z$_\odot$ = 0.019. Contrastingly, \citet{kudryavtseva12} provided an age estimate of $2 \pm 0.1$ Myr with assumed solar metallicity of Z$_\odot$ = 0.015, which is more consistent with the estimates from the upper CMD.
{\citet{harayama08} also estimated an age of 0.5-1.0 Myr for the PMS stars in NGC 3603.}
While this issue of PMS stars is interesting in itself, it is beyond the scope of this paper and in the current context it makes more sense to compare the SSP results with those from the upper CMD, as the lower CMD contributes negligible amount to the integrated output light of the stellar population.

\begin{figure*}[]
\centering
\begin{subfigure}{.5\textwidth}
  \centering
  \includegraphics[width=\linewidth]{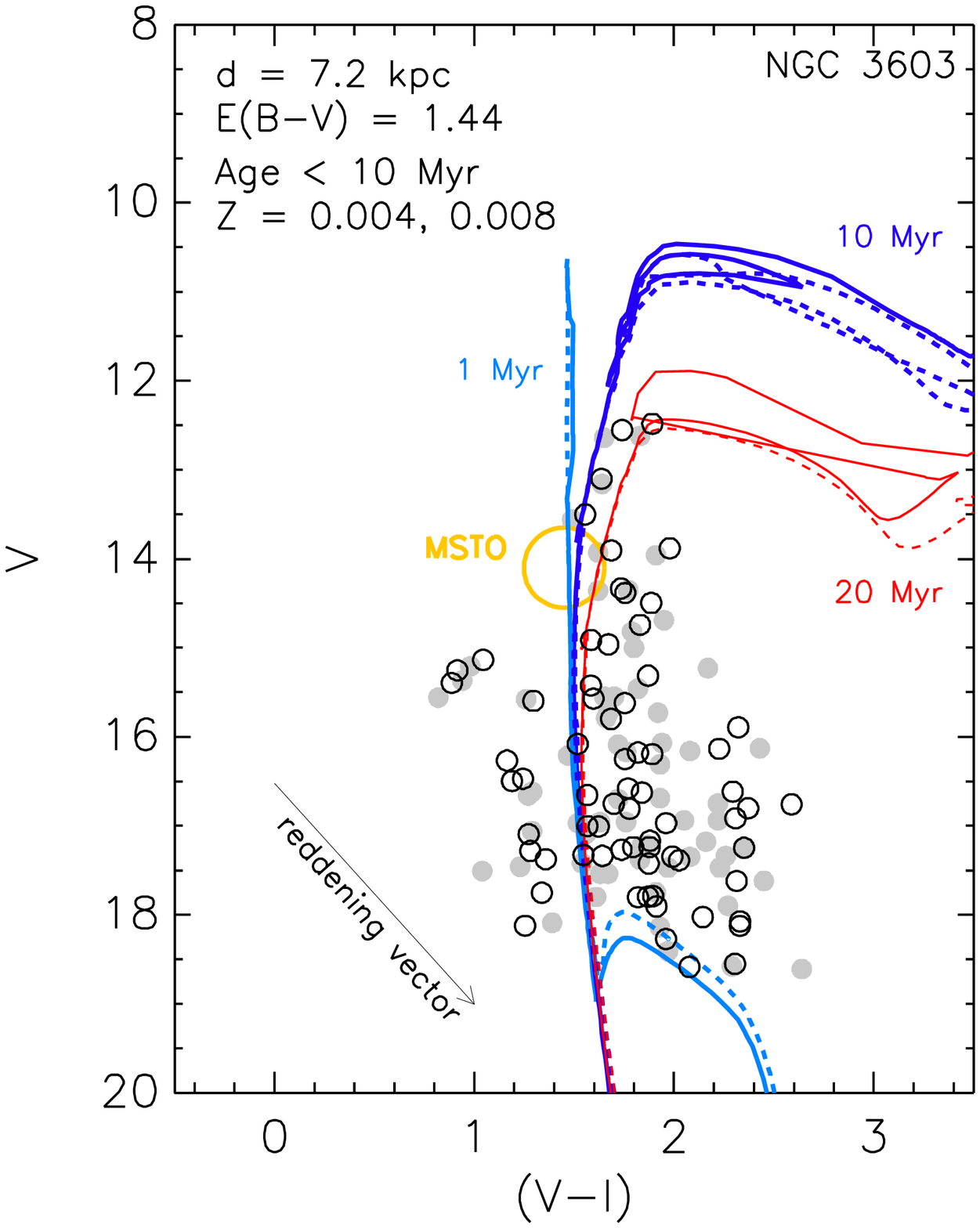}
\end{subfigure}%
\begin{subfigure}{.5\textwidth}
  \centering
  \includegraphics[width=\linewidth]{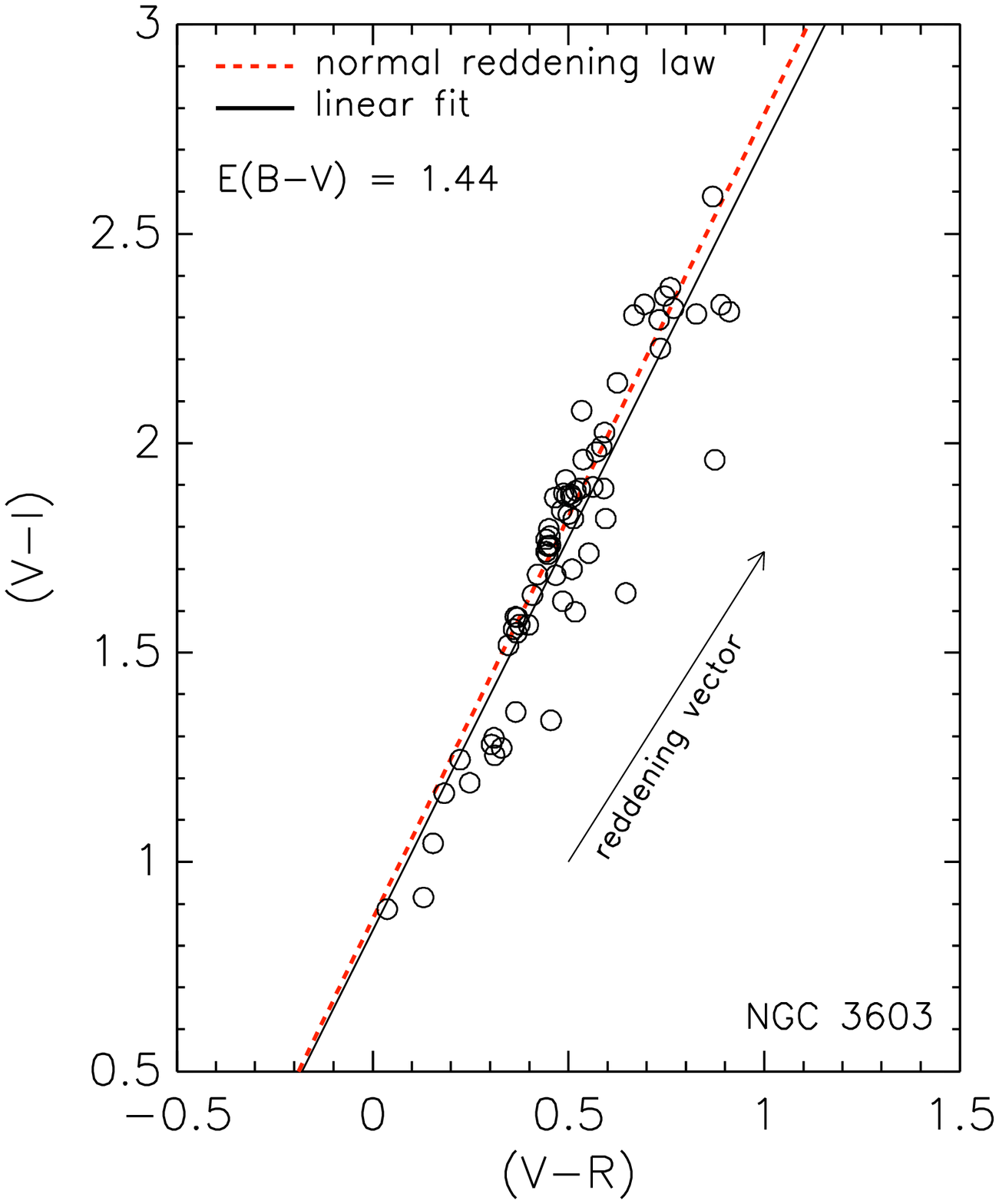}
\end{subfigure}
\caption{\textit{(Left:)} color-magnitude diagram of NGC 3603, constructed using synthetic photometry (open black circles). Filled grey circles represent the WEBDA catalog magnitudes \citep{sagar01}, shown for comparison.
Isochrones of $Z$ = 0.004, log $t$ = 6.0 (1 Myr, cyan) and 7.0 (10 Myr, blue) are plotted with solid lines, while those of the same ages at $Z$ = 0.008 with dashed lines. Non-fitting isochrones of 20 Myr are also plotted (red).
{The position of the main sequence turn-off point (MSTO) is indicated {with a yellow circle}.}
\textit{(Right:)} two-color diagram of NGC 3603, also constructed using synthetic photometry (open black circles). Solid black line indicates a linear regression, while dashed red line indicates the normal interstellar extinction law. Following \citet{sagar01}, color excess ratios of $E(V-R)/E(B-V) = 0.65$ \citep{alcala88} and $E(V-I)/E(B-V) = 1.25$ \citep{dean78} are used for the normal extinction law. The directions of the normal reddening vectors are shown with arrows in both diagrams.
}
\label{cmdtcd}
\end{figure*}

We demonstrate that our integral field spectroscopy is able to reproduce the photometry and the cluster CMD, on which a procedure of isochrone fitting was performed to estimate the cluster age. The isochrones used are from the PARSEC database \citep{bressan12}.
Figure \ref{cmdtcd} shows the CMD and two-color diagram commonly used in resolved stellar population studies, generated using the synthetic magnitudes calculated from the individual star spectra. 
{The diagrams contain 68 individual stars within the field of view.}
The CMD of NGC 3603 shows a well-defined main sequence, with a contamination of foreground stars suffering less reddening compared to the cluster stars, on the blue side of the main sequence at $(V-I) < 1.4$. 

{In the CMD (Figure~\ref{cmdtcd}), the cluster main sequence turn-off  (MSTO) point is indicated. 
It is the position where, starting with the most massive stars within the cluster, stars begin to depart from the zero-age main sequence (ZAMS; here represented by the 1 Myr isochrone) and moving towards the (super)giant branch. 
{Therefore, the MSTO essentially signifies the position of the bluest and most massive star still in the main sequence.} In an SSP with zero extinction, the distribution of the position of the stars would closely follow the isochrone shape. In reality, in most clusters the observed main sequence is broadened as stars are spread redward of the isochrone along the reddening vector. NGC 3603 suffers from this effect of differential reddening \citep{sagar01,melena08} and displays such broadened main sequence.
Thus, the position of a star in the CMD does not necessarily reflect its actual position on the isochrone. Its actual, dereddened, position lies in the isochrone in the direction of the reverse of the reddening vector. The star's position cannot be bluer than the isochrone.
In the case of NGC 3603, the stars above the MSTO could belong to the 10 Myr isochrone or younger, but not to the older ones (e.g. 20 Myr) as the isochrone would be too red compared to those stars. Therefore, it is unlikely that the cluster is aged more than 10 Myr.
}

{Thus, in the CMD} the photometry can be well fit with isochrones of ages less than 10 Myr for both $Z$ = 0.004 and $Z$ = 0.008, while fixing the values of distance and reddening as given in \citet{sagar01}, {i.e., $d$ = 7.2 kpc, $E(B-V)$ = 1.44}. The adopted values of distance and reddening differ by only about 10\% with those in the other references
{(e.g. $d$ = 6.3 kpc, $E(B-V)$ = 1.48, \citealt{pandey00}; $d$ = 7.6 kpc, $E(B-V)$ = 1.39, \citealt{melena08}.
It was also noted by \citet{melena08} that there is very good agreement on the derived apparent distance modulus of NGC 3603 in the literature, with the uncertainty on the physical distance stemming mainly from the extinction uncertainty.
}


The two-color diagram also shows that the interstellar extinction law towards the direction of NGC 3603 is quite normal, in agreement with the result of \citet{sagar01}.
Varying the metallicity of the isochrone does not affect the isochrone fitting as the isochrone shape is only weakly sensitive to metallicity variation. However, isochrones are very age-sensitive and in this case the age solution is limited up to 10 Myr, as beyond this age the cluster turn-off point would move too far downward.

This upper limit of 10 Myr does not contradict our spectroscopic age estimate of {4-6} Myr. Furthermore, the photometric ages found in the literature agree reasonably well with the spectroscopic age estimate. 
{It has been reported that the age of the stars in the outer part of the cluster reach up to 4-5 Myr \citep{melnick89,pandey00,sagar01,sung04,melena08}. Indeed the analysis in the current work corresponds more closely to this outer region.}
Table~\ref{table:1} presents our estimates of age and metallicity compared to a number of references. 
These indicate that the age estimates derived from resolved and unresolved stellar populations are broadly consistent. We note that there may be present a number of caveats affecting the CMD and isochrone fitting such as differential reddening, binarity, sequential star formation, field star contamination and selections of stars included in analysis, as well as different isochrone prescriptions. 
On the other hand, the SSP models used in the spectroscopic analysis are also subject to different prescriptions used to generate them, as demonstrated in Figure \ref{haewcomp}. We note that our field of view does not contain the whole extent of the NGC 3603 cluster and the associated nebula\footnote{A mosaic observation was requested but not obtained due to limitations in the science verification run of the instrument.} (see Figure \ref{figrgb}), which may affect the derived properties.
{Finally, due the low number of stars in our CMD, the result is not very constraining compared to the other CMD results in the literature which converge at around 1-5 Myr.}

We point out that direct comparison between photometry and spectroscopy in the context of metallicity determination was not conclusive due to the nature of the isochrones.
The methods utilizing isochrone or SED fitting are not very sensitive to metallicity variations, therefore generally are considered inferior compared to the strong line method in determining metallicity. However, in the case of NGC 3603 we confirm that the metallicity values obtained from the integrated spectrum can be accommodated by isochrone fitting.
In Table~\ref{table:1} the reference metallicities are assumed or derived spectroscopically.

\begin{table*}
\caption{Age and metallicity estimates obtained in this study, compared to previous works.}             
\label{table:1}      
\centering                          
\begin{tabular}{c c c c}        
\hline\hline                 
Reference & Age (Myr) & $Z$ & Notes \\    
\hline                        
\multicolumn{4}{c}{{\textit{Unresolved stellar population}}} \\
   This work & {$\sim$ 4-6}  & 0.004 &  Age from H$\alpha$EW, $Z$ from strong lines \\   
   This work & {$\sim$ 4}  & 0.008 & Age and $Z$ from SED fitting \\
\multicolumn{4}{c}{{\textit{Resolved stellar population}}} \\
   \citet{melnick89} & $\sim2.5 \pm 2$ & 0.007 & Age from CMD, $Z$ from strong lines \\ 
   \citet{pandey00} & $\lesssim 1$ & ---    &  Age from CMD \\
   \citet{sagar01} & $3\pm2$ & 0.020 & Age from CMD, solar $Z$ assumed \\
{   \citet{sung04} } & $\lesssim5$ & --- & {Age from CMD} \\
   {\citet{harayama08}} & $\sim 2.5$ & 0.020 &  {Age from CMD, solar $Z$ assumed} \\
   \citet{melena08} & $\lesssim4$ & 0.020 & Age from CMD, solar $Z$ assumed \\
      {This work} & $\lesssim 10$ & --- & {Age from CMD} \\
\hline                                   
\end{tabular}
\end{table*}

\bigskip

\section{Summary}

We have obtained integral field spectroscopy of NGC 3603 using VLT/MUSE, from which an analysis to determine age and metallicity was performed with the aim of {testing the validity of} SSP models. 
{The integrated spectrum of NGC 3603 was analysed by utilizing the strong-line method, together with SSP model comparisons to both the \haew, and the overall SED.
The resulting age and metallicity were then compared to those derived from an independent method utilizing the color-magnitude diagram of resolved stellar population {from this very dataset}, and also to other age and metallicity estimates obtained from the literature.}

From the integrated spectrum the age of NGC 3603 was derived as around {4-6} Myr and with metallicity between $Z$ = 0.004 and $Z$ = 0.008. {The uncertainty in the age estimate originates from the different results when comparing the observed \haew$ $ to the different SSP models (BPASS/GALEV/Starburst99). In the case of SED fitting, using different SSP models similarly alters the age estimate although this also results in large variations in metallicity and fraction of contamination from older populations. Overall this illustrates the notion that using different SSP models may yield different results in the interpretation of stellar population properties.
}

{The 4-6 Myr age estimate from the integrated spectrum} agrees reasonably well with the estimates from photometry ({$\sim$ 1-5 Myr}), and suggests the consistency between the two independent methods. This pathfinder study provides a promising start towards a more systematic and comprehensive investigation in the effort of {testing} SSP models for young stellar populations. In this context, IFS offers a unique way to perform such investigation due to the extended nature of the ionized gas component in young stellar populations. 
{MUSE as the foremost IFS instrument currently available offers the exciting capability of exploring both the stellar content and nebular component of young star clusters simultaneously.}
{For systems located close enough as to have a good spatial resolution, MUSE allows both resolved stellar population and integrated population analyses.}
Wide-field IFS observations of young stellar clusters with a significantly larger sample are desired to further advance this study{, opening the possibility of a new field of analysis.}
{We encourage the community to undertake efforts such as the one presented here in comparing resolved and unresolved stellar populations. This would help in refining SSP analysis techniques, which is ultimately vital for our understanding of extragalactic systems.}


\begin{acknowledgements}
The authors thank the anonymous referee whose valuable comments and suggestions helped improve the paper.
We also wish to thank Patricia S\'anchez-Bl\'azquez, Mamoru Doi, and Nobuo Arimoto for reading and providing useful comments for the manuscript.
Support for HK, LG, and MH is provided by the Ministry of Economy, Development, and Tourism's Millennium Science Initiative through grant IC120009, awarded to The Millennium Institute of Astrophysics, MAS. HK and LG acknowledge support by CONICYT through FONDECYT grants 3140563 and 3140566, respectively.
TK acknowledges support through the Sofja Kovalevskaja Award to P. Schady from the Alexander von Humboldt Foundation of Germany.
We acknowledge great help provided by the MUSE Science Verification Team.
Based on observations collected at the European Organisation for Astronomical Research in the Southern Hemisphere under ESO programme 60.A-9344. 
This research has made use of the WEBDA database, operated at the Department of Theoretical Physics and Astrophysics of the Masaryk University.
\end{acknowledgements}


\end{document}